\documentclass[11pt,letter]{article}

\usepackage{special,times,graphicx}
\usepackage{subfigure}

\usepackage{booktabs}

\usepackage{subfigure}
\usepackage{caption}
\usepackage{url}
\usepackage{color}
\usepackage{float}

\begin{document}

\newlength{\bibitemsep}\setlength{\bibitemsep}{.2\baselineskip plus .05\baselineskip minus .00\baselineskip}
\newlength{\bibparskip}\setlength{\bibparskip}{0pt}
\let\oldthebibliography\thebibliography
\renewcommand\thebibliography[1]{%
  \oldthebibliography{#1}%
  \setlength{\parskip}{\bibitemsep}%
  \setlength{\itemsep}{\bibparskip}%
}

\title{GeoCoV19: A Dataset of Hundreds of Millions of Multilingual COVID-19 Tweets with Location Information}

\author{Umair Qazi, Muhammad Imran, Ferda Ofli\\\{uqazi, mimran, fofli\}@hbku.edu.qa\\Qatar Computing Research Institute, Hamad Bin Khalifa University, Doha, Qatar}

\maketitle
\begin{abstract}
The past several years have witnessed a huge surge in the use of social media platforms during mass convergence events such as health emergencies, natural or human-induced disasters. These non-traditional data sources are becoming vital for disease forecasts and surveillance when preparing for epidemic and pandemic outbreaks. In this paper, we present GeoCoV19, a large-scale Twitter dataset containing more than 524 million multilingual tweets posted over a period of 90 days since February 1, 2020. Moreover, we employ a gazetteer-based approach to infer the geolocation of tweets. We postulate that this large-scale, multilingual, geolocated social media data can empower the research communities to evaluate how societies are collectively coping with this unprecedented global crisis as well as to develop computational methods to address challenges such as identifying fake news, understanding communities' knowledge gaps, building disease forecast and surveillance models, among others.
\end{abstract}

\section{Introduction}
\label{sec:intro}
Social media platforms such as Twitter receive an overwhelming amount of messages during emergency events, including disease outbreaks, natural and human-induced disasters. In particular, the information shared on Twitter during disease outbreaks is a goldmine for the field of epidemiology. Research studies show that Twitter provides timely access to health-related data about chronic disease, outbreaks, and epidemics~\cite{broniatowski2013national}. In this paper, we present GeoCoV19, a large-scale Twitter dataset about the COVID-19 pandemic. Coronavirus disease 2019 or COVID-19 is an infectious disease that was first identified in December 2019 and has spread globally since then. The World Health Organization (WHO) declared the COVID-19 outbreak a pandemic on March 11, 2020. As of May 6, 2020, more than 263K fatalities have been recorded, and more than 3.8 million people have been infected globally. Twitter has seen a massive surge in the daily traffic amid COVID-19. People try to connect with their family and friends, look for the latest information about the pandemic, report their symptoms, and ask questions. Moreover, many conspiracies, rumors, and misinformation began to surface on social media, e.g., drinking bleach can cure it, Bill Gates is behind it, etc. Furthermore, both benefits and unanticipated consequences of lockdowns and closure of businesses around the globe and social distancing are among the top topics on social media. 

In this dataset collection, we aimed at covering several different perspectives related to the COVID-19 pandemic ranging from social distancing to food scarcity and symptoms to treatments and shortage of supplies and masks. The dataset contains more than 524 million multilingual tweets collected over a period of 90 days starting from February 1, 2020 till May 1, 2020, using hundreds of multilingual hashtags and keywords. Various public health and disease surveillance applications that use social media rely on broad coverage of location information. However, only a limited number of tweets contain geolocation information (i.e., 1-3\%). To increase the geographical coverage of the dataset, we employ a gazetteer-based approach that leverages various meta-data information from Twitter messages to infer their geolocation. The dataset contains around 378K geotagged tweets with GPS coordinates and 5.4 million tweets with \textit{place} information. However, with the help of the proposed geolocation inference approach, we extracted additional geolocation information for 297 million tweets using the location information from the user profile and for 452 million tweets using toponyms in tweet content. Consequently, our dataset contains around 491 million tweets with at least one type of geolocation information, which constitutes 94\% of the entire dataset.
Overall, there are 43 million unique users in the dataset, which includes around 209K users who have verified Twitter accounts. In terms of its multilingualism, the dataset covers 62 international languages. Among top languages, there are approximately 348 million English, 65 Spanish, 9.3 million Italian, and 5.5 million Arabic tweets.

The rest of the paper is organized as follows. In the next section, we describe COVID-19 related datasets, specifically from Twitter. In Section~\ref{sec:data_colection}, we provide the details of data collection. Section~\ref{sec:geo_infer} explains the gazetteer approach followed for geolocation inference. Section~\ref{sec:data_desc} presents various statistics about the dataset followed by a section on research implications. Finally, we conclude the paper in Section~\ref{sec:conclusion}.

\section{Related Work}
\label{related_work}



The number of preliminary studies on analyzing social media data related to the COVID-19 pandemic has been increasing almost on a daily basis. Some researchers analyze human behavior and reactions to the spread of COVID-19 in the online world~\cite{li2020characterizing,abd2020top,rashid2020covidsens,schild2020go}, or investigate conspiracy theories and social activism~\cite{shahsavari2020conspiracy,ferrara2020covid}. Many others collect and share large-scale datasets publicly to enable further research on the topic. Some datasets are restricted to a single language such as English~\cite{lamsal2020coronasentiment,lamsal2020coronatweets} or Arabic~\cite{alqurashi2020large,haouari2020arcov} while some others contain multiple languages~\cite{chen2020covid,banda2020large,kerchner2020coronavirus}. The data collection periods also vary between these datasets. Among all, \cite{banda2020large} stands out as one of the long-running collections with the largest amount of tweets (i.e., 250 million), also thanks to its multilingual nature. However, these datasets mostly contain only the raw content obtained from Twitter except that \cite{lamsal2020coronasentiment,lamsal2020coronatweets} associate a sentiment score with each tweet, while \cite{haouari2020arcov} provides tweet propagation networks. On the other hand, our dataset enriches the raw tweet content with additional geolocation information, which is otherwise almost non-existent. Furthermore, with more than 524 million tweets, our dataset is more than twice as large as the largest dataset available to date.

\begin{figure}[!h]
\centering
\includegraphics[width=\textwidth]{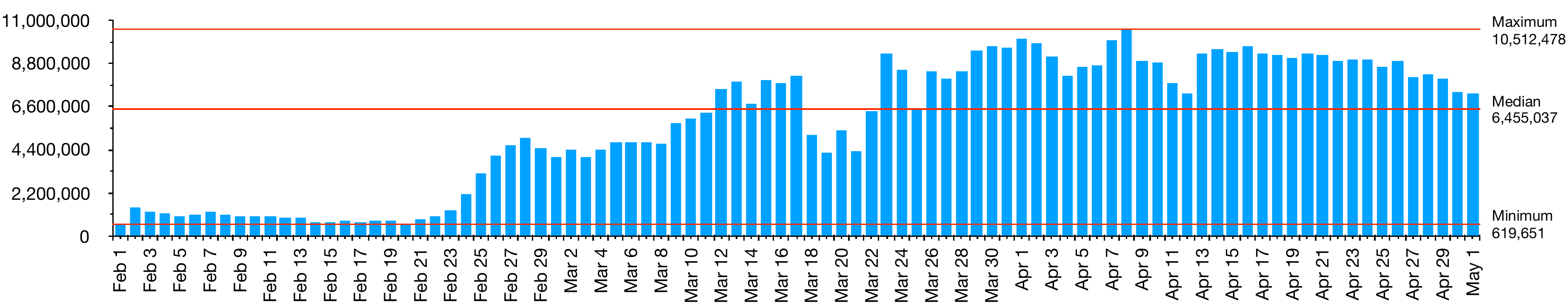}
\caption{Daily distribution of 524,353,432 tweets from February 1, 2020 to May 1, 2020.}
\label{fig:data_daily_dist}
\end{figure}

\section{Data Collection}
\label{sec:data_colection}
Twitter offers different APIs for data collection. We use the Twitter Streaming API through which one can collect data using hashtags, keywords, or geographical bounding boxes. We used the AIDR system~\cite{imran2014aidr}\footnote{\url{http://aidr.qcri.org}} for the data collection through hundreds of different multilingual hashtags and keywords related to the COVID-19 pandemic. The data collection started on February 1, 2020 using trending hashtags such as \#covid19, \#coronavirus, \#covid\_19, and new hashtags were added as they emerged in the later days. As of May 1, the total number of hashtags/keywords is 803. Although Twitter offers filtered streams for specific languages, we do not filter by any language; thus, our data is multilingual. As of May 1, 2020, we have collected 524,353,432 tweets in total. Our data collection still continues, however all the statistics and analyses presented in this paper are based on tweets collected till May 1, 2020. 

Figure~\ref{fig:data_daily_dist} depicts the daily ingested volume of tweets during the 90 days of data collection (February 1 to May 1, 2020). The data does not show any gaps. However, the daily volume in the first two weeks is lower than the later weeks. On average, ${\sim}$1 million tweets were ingested daily during the first two weeks. Whereas the average daily volume substantially increased during the later weeks, i.e., around 8 million tweets per day. The maximum number of tweets recorded in one day is ${\sim}$10.5 million on April 8.

We make the dataset publicly available for research and non-profit uses. Adhering to Twitter data redistribution policies, we cannot share full tweets content. Therefore, we only share tweet ids and user ids, along with high-level geolocation information for each tweet. Moreover, we develop and provide a tool to rehydrate full tweet content using the shared tweet ids. Weekly data files and all the hashtags and keywords used for the collection are available at: \url{https://crisisNLP.qcri.org/covid19}

\section{COVID-19 Tweets Geolocation Inference}
\label{sec:geo_infer}
\subsection{Problem} 
The problem of geolocation inference on Twitter data can be divided into two sub-problems: (i) user geolocation, i.e., determining the location of a user, and (ii) tweet geolocation, i.e., determining the location of a tweet. The user geolocation inference problem can be further categorized as (a) user home location inference and (b) user current location inference. Similarly, the tweet geolocation inference problem can be further divided as (a) determining the origin of a tweet (i.e., the location where the tweet is posted from) and (b) extracting and disambiguating the mentioned locations in a tweet. Despite these subtle distinctions, many research studies map all of these problems into one, which is the tweet geolocation inference. Although the dataset described in this paper can be used to address each of these sub-problems, hereinafter we refer all of them as the tweet geolocation inference problem.

\subsection{Data for inference}
Although Twitter provides an option to enable the exact geolocation service on mobile devices, only 1\% of the tweets contain GPS coordinates, i.e., \textit{latitude} and \textit{longitude}. An alternative option to infer geolocation of a tweet is to use the \textit{place} field---an optional field that the user fills by selecting one of the Twitter-suggested locations. This place tag associates a tweet with a location referring to a city, an area, or a famous point-of-interest such as a building or a restaurant. Twitter provides the place field data in a JSON dictionary that contains several types of information including city or country name, and a bounding box representing the specified location. When present, the \textit{place} field indicates that the tweet is associated with a place but may not be originating necessarily from that place.

Another widely used meta-information for geolocation inference is \textit{user location}, which is provided by the user in their profile. This is an optional free-form text field where values such as United States, USA, US, NYC, Earth, Moon are all valid entries but not necessarily valid location names. Followers and friends networks can also help determine the home location of a user. Alternatively, tweet content, i.e., the actual tweet message, can have one or more location mentions, which can be extracted for the location inference task.

\subsection{Toponym extraction and standardization}
We use four types of information from a tweet: (i) geo-coordinates (latitude, longitude), (ii) place field (JSON dictionary), (iii) user location field (text), and (iv) tweet content (text). To infer locations from textual content, we tried several publicly available Named-Entity Recognition systems including NLTK\footnote{\url{https://www.nltk.org}}, a recent version of Stanford NER~\cite{finkel2005incorporating}, and a system by~\cite{ritter2011named}. However, due to either unsatisfactory results (e.g., most of them need proper capitalization to detect location names) or not enough computational speed to deal with our large dataset, we were unable to employ these tools in our study. Instead, we have adopted a gazetteer-based approach and used Nominatim, which is a search engine for OpenStreetMap (OSM) data.\footnote{OSM is a collaborative project and provides free geographic data such as street maps.} Specifically, we used the Nominatim tool to perform two operations: (i) geocoding---the process of obtaining geo-coordinates from a location name, and (ii) reverse geocoding---the process of obtaining a location name from geo-coordinates. The official online Nominatim service does not permit more than 60 calls/minute, and hence, is not suitable for our purpose to make millions of calls in a reasonable time period. Therefore, we set up six local Nominatim servers on our infrastructure and tuned each server to handle 4,000 calls/second. Next, we describe in detail how we determine the geolocation of a tweet.

\begin{itemize}
    \item {\bf Geo-coordinates:} Although geo-coordinates can be directly used for mapping, we perform reverse geocoding to obtain additional location details, e.g., street, city, or country the coordinates belong to. For this purpose, we use Nominatim's \textit{reverse} API endpoint to extract city, county, state, and country information from the Nominatim response if available.

    \item {\bf Place:} The place field data contains several location attributes. We use the \textit{full name} attribute, which represents the name of a location in textual form, and query Nominatim's \textit{search} API endpoint to extract city, county, state, and country names from the Nominatim response if available.

    \item {\bf User location:} The user location field is a free-form text entered by the user, and hence, is often noisy as the users can type anything, e.g., ``California, USA", ``Earth", ``Mars", and ``home". We follow the procedure explained below for toponym extraction from text using this field. The extracted toponyms are then used to obtain geolocation information including country, state, county, and city from Nominatim.

    \item {\bf Tweet content:} Tweet content represents the actual tweet in text format. To extract location mentions inside tweet text, we follow the toponym extraction from text procedure described below. The extracted toponyms are then used to obtain geolocation information at various levels such as country, state, county, city, and GPS-coordinates (if available).
\end{itemize}

\noindent{\bf Toponym extraction from text.} Given a piece of text, either from the user location field or the actual tweet content, we follow the below steps to extract geolocation information from it:

\begin{enumerate}
    \item {\bf Preprocessing:} Replace all URLs, RT, user mentions (starting with @ sign), numbers, and other special characters with a ``\textless noise\textgreater"  token. All remaining words are lowercased.
    \item {\bf Candidate generation:} All remaining single words (i.e., uni-grams) and their bi-grams are considered toponym candidates after dropping bi-grams that contain the \textless noise\textgreater~token. The bi-grams ensure that locations with more than one term such as ``new york" are not missed.
    
    \item {\bf Non-toponyms pruning:} All candidates (uni-grams and bi-grams) are checked (i) whether they are stop-words and (ii) whether they appear in an index consisting of 3,174,209 location names taken from Kaggle.\footnote{\url{https://www.kaggle.com/max-mind/world-cities-database}} This O(1) operation removes all stop-words as well as those that are not found in the location index. Removing stop-words after generating bi-grams allows us to account for toponyms such as ``new york".
    \item {\bf Nominatim:} The remaining candidates are used to query Nominatim. Specifically, we make a Nominatim \textit{search} call for each candidate. For valid toponym queries, Nominatim returns a non-empty response.
    \item {\bf Majority voting (optional):} In case there are multiple toponyms with non-empty Nominatim response, we resolve them to one location using majority voting based on country information.
    
\end{enumerate}
\vspace{-0.3cm}
\subsection{Evaluation}\vspace{-0.2cm}
Here we evaluate the accuracy of the toponym extraction approach described above. The two fields from which we extract toponyms include user location and tweet content. As described earlier, the extracted toponyms are used to obtain geolocation information at different granularity levels such as country, state, county, and city. For this purpose, we take a random sample of 5,000 tweets with exact GPS-coordinates and for which the geolocation information (i.e., country, state, county, and city) is derived from the user location and tweet content fields. The evaluation aims to determine if the derived geolocation information at different granularity levels (i.e., country, state, county, and city) matches with the geolocation information of the tweets with GPS-coordinates. 

\begin{table}[h]
\centering
\caption{Accuracy of the toponym extraction approach at different granularity levels}
\label{tbl:eval}
\begin{tabular}{lcccc}
\toprule
{\bf Tweet field} & {\bf Country} & {\bf State} & {\bf County} & {\bf City}\\
\midrule
User location & 0.86 & 0.62 & 0.34 & 0.27\\ 
Tweet content & 0.75 & 0.48 & 0.29 & 0.23\\ 
\bottomrule
\end{tabular}
\end{table}

Table~\ref{tbl:eval} shows evaluation results in terms of accuracy for four geolocation granularity levels. We see that the lower the granularity (i.e., higher administrative level), the better the accuracy scores. At the country level, for both user location and tweet content fields, the accuracy scores are plausible at 0.86 for user location and 0.75 for tweet content. At the state level, the user location field seems to yield better results than the tweet content. However, for both county and city levels, the accuracy scores are low.
\vspace{-0.3cm}\section{Data Description}\vspace{-0.2cm}
\label{sec:data_desc}
This section describes various details of the dataset such as geographical coverage, users, and multilingualism. 
\smallskip

\begin{figure}[t]
\centering     
\subfigure[Organic geotagged tweets]{\label{fig:geo_map}\includegraphics[width=0.32\textwidth]{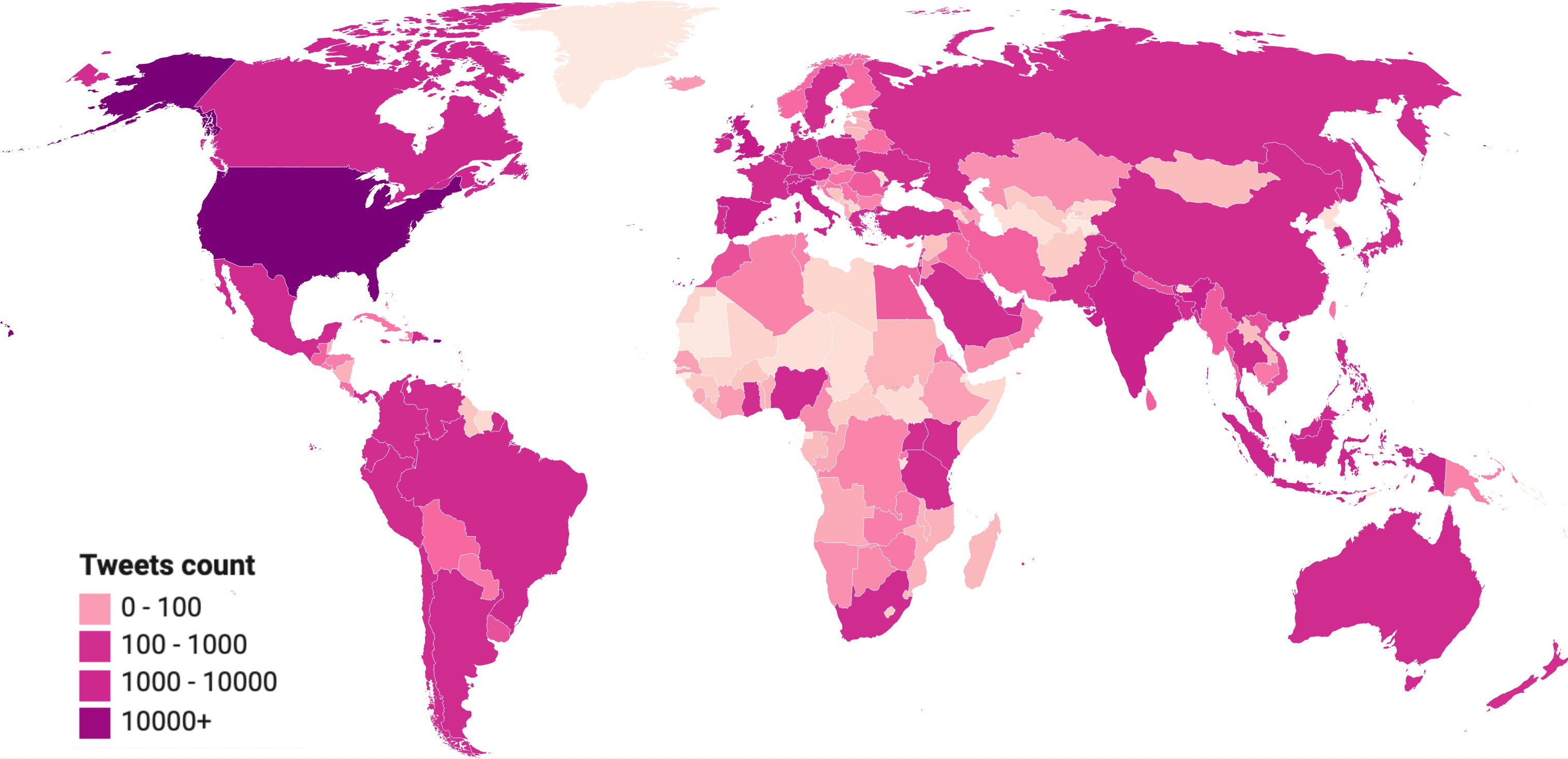}}
\subfigure[User distribution]{\label{fig:users_map}\includegraphics[width=0.32\textwidth]{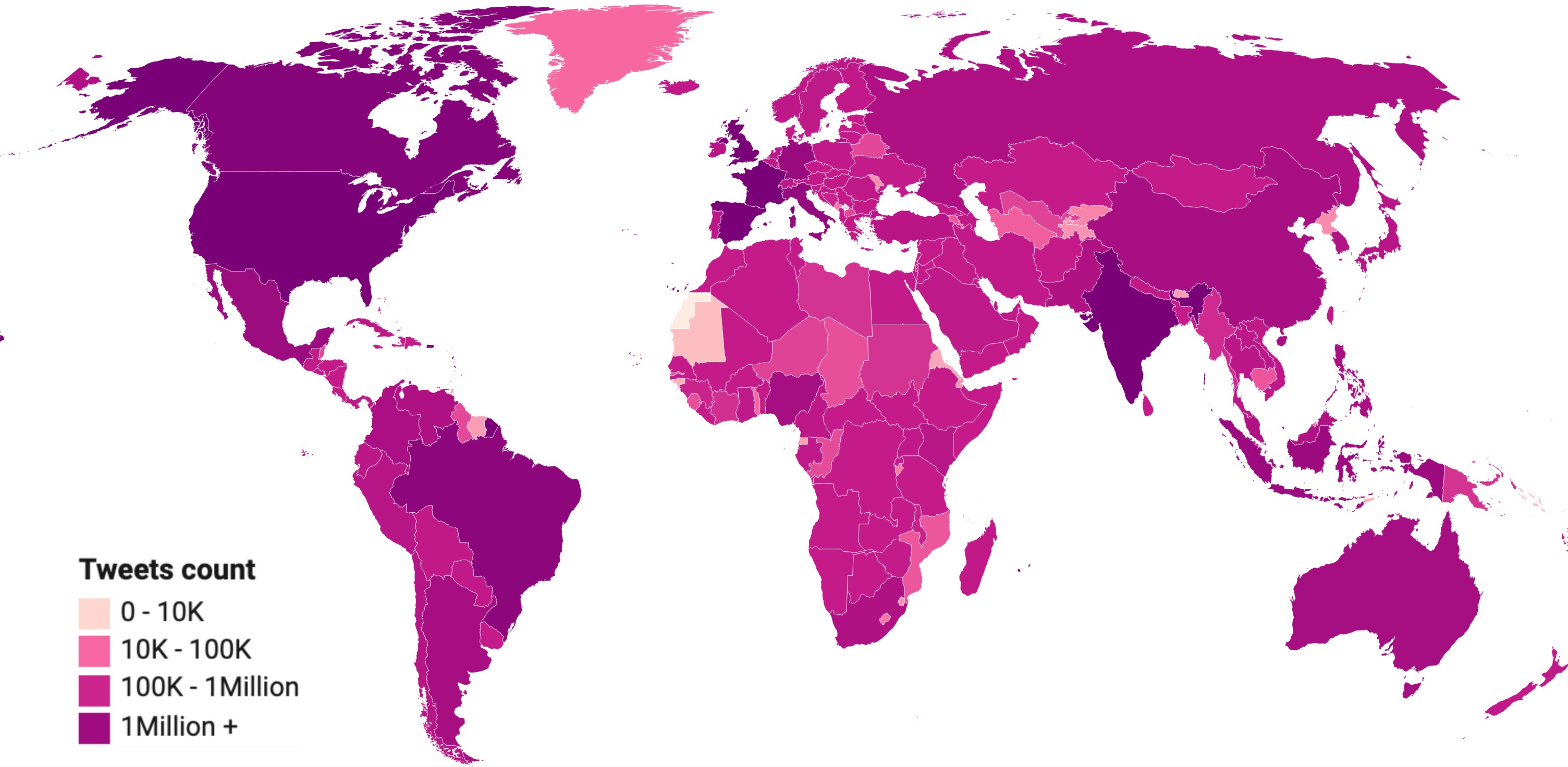}}
\subfigure[Most-mentioned locations]{\label{fig:most_mentioned_map}\includegraphics[width=0.32\textwidth]{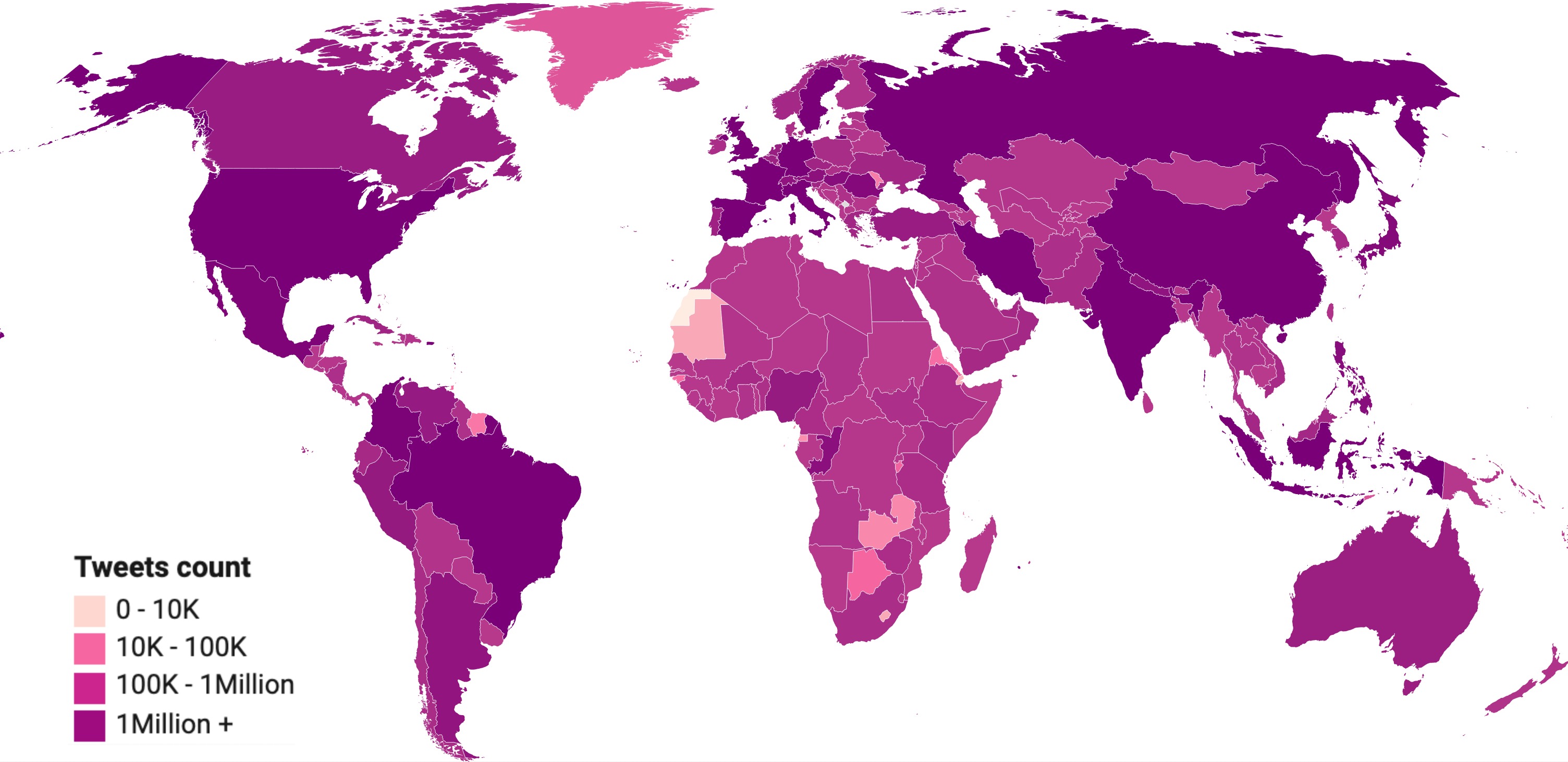}}\vspace{-0.3cm}
\caption{Geographic coverage of the dataset\vspace{-0.2cm}}
\end{figure}

\noindent{\bf Geolocated tweets and their geographical coverage}\\
Table~\ref{tbl:tweets_geo} shows the details of tweets with different location fields and their quantities in the dataset. The dataset contains 378,772 geotagged tweets (with latitude, longitude) for which the geolocation information is received from users' devices (GPS), and it is the most trusted and accurate geolocation information. Around 5.4 million tweets are tagged with \textit{place} information, which is the second most reliable geolocation information. Figure~\ref{fig:geo_map} shows the combined geographical distribution of these geotagged tweets on a world map. 
The toponym extraction approach described in Section~\ref{sec:geo_infer} was employed to identify geolocation information from the \textit{user location} and \textit{tweet content} fields. Using this approach, we were able to find geolocation information for around 297 million tweets using the \textit{user location} field and around 452 million tweets using the \textit{tweet content} field. Figure~\ref{fig:users_map} shows the geographical distribution of the users for which the geolocation information is derived from the user location field. Figure~\ref{fig:most_mentioned_map} shows the geographical distribution of the most mentioned locations extracted from the tweet content.

\begin{figure}[ht]
\centering     
\subfigure[Top-15 countries' daily distribution]{\label{fig:cc_daily_dist}\includegraphics[width=0.815\columnwidth]{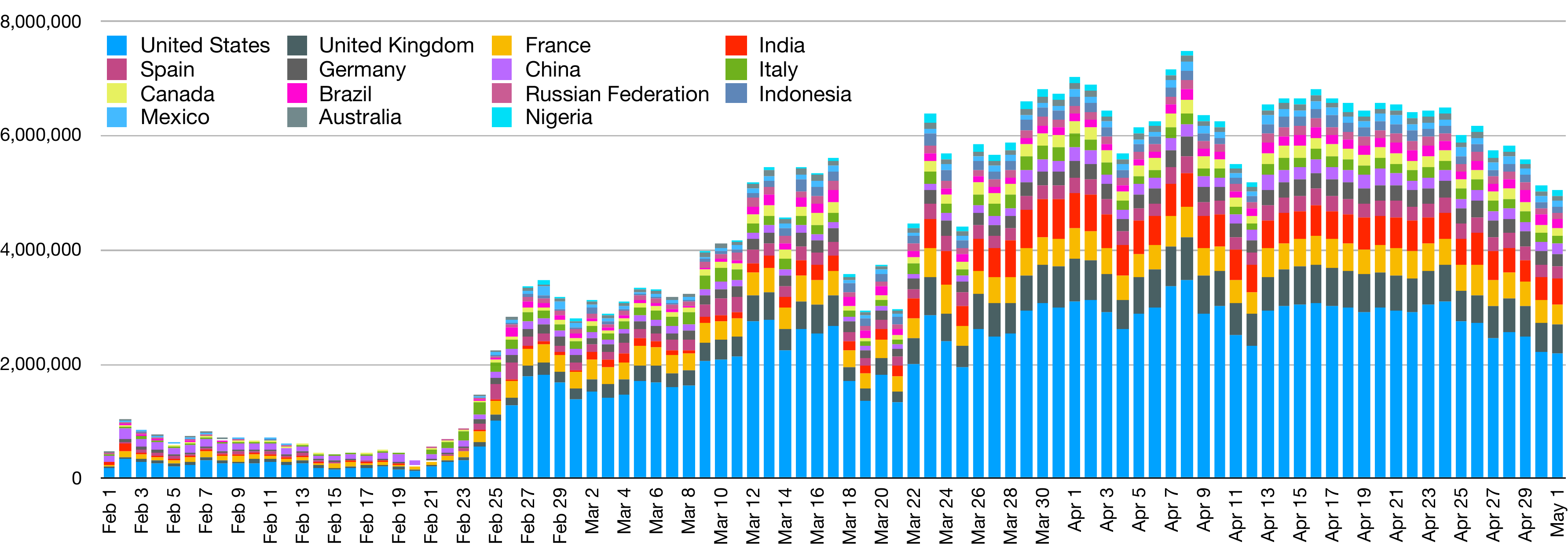}}
\subfigure[Top-15 cities' daily distribution]{\label{fig:cities_daily_dist}\includegraphics[width=0.815\columnwidth]{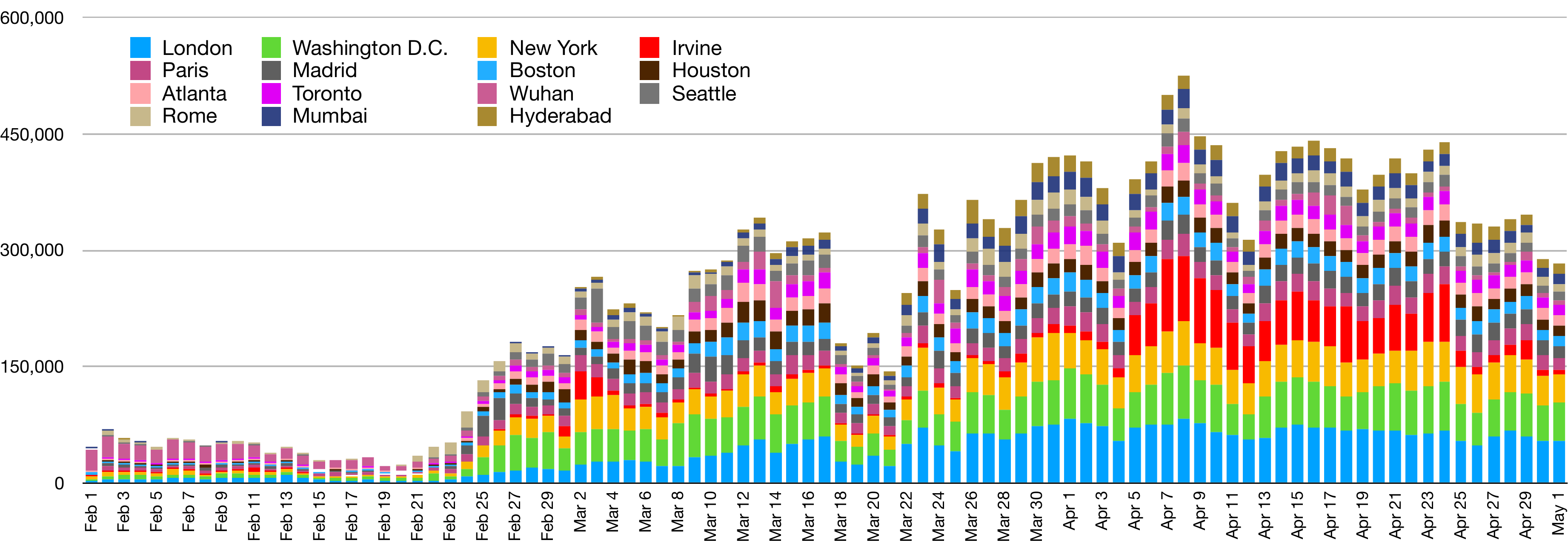}}\vspace{-0.4cm}
\caption{Countries and cities distributions across days\vspace{-0.1cm}}
\end{figure}

The dataset covers 218 countries and 47,328 unique cities worldwide, and several types of amenities, such as hospitals, parks, and schools. The countries also include islands such as the British Virgin Islands and the Solomon Islands. Table~\ref{tbl:country_vol_bucket} shows countries with tweet volume breakdown. There are 9 countries with more than 10 million tweets, 40 with more than 1 million tweets, and 28 with more than 500K tweets. Table~\ref{tbl:cities_vol_bucket} shows a similar breakdown for cities. Figure~\ref{fig:cc_daily_dist} shows the daily distribution of top-15 countries and Figure~\ref{fig:cities_daily_dist} depicts the daily distribution of top-15 cities where London, Washington D.C., and New York are the top-3 cities. 

\begin{figure}
\begin{minipage}[b]{1\textwidth}
\centering
\subfigure[USA: Organic geotagged tweets]{\label{fig:usa_geo_map}\includegraphics[width=0.32\textwidth]{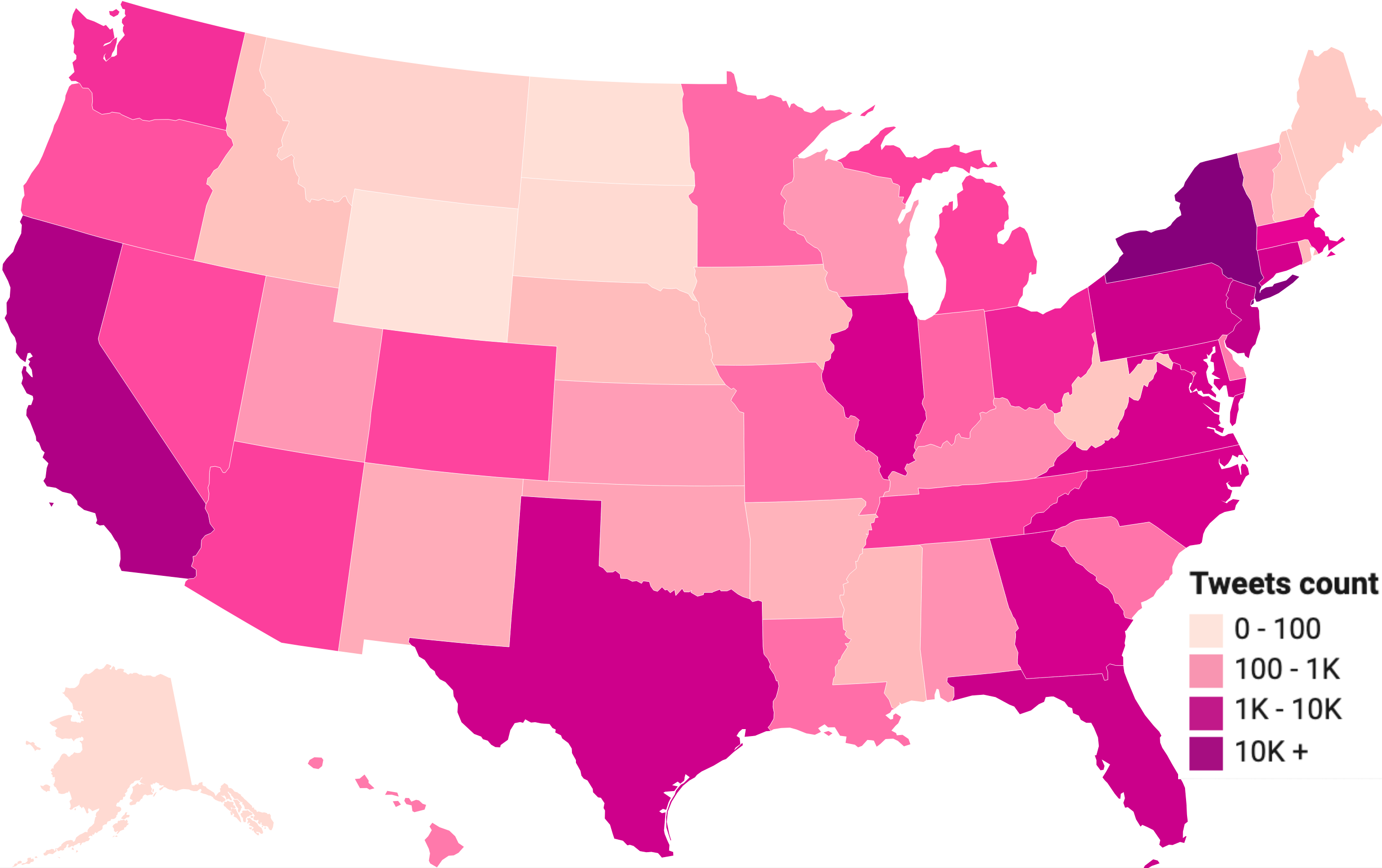}}
\subfigure[USA: User distribution]{\label{fig:usa_users_map}\includegraphics[width=0.32\textwidth]{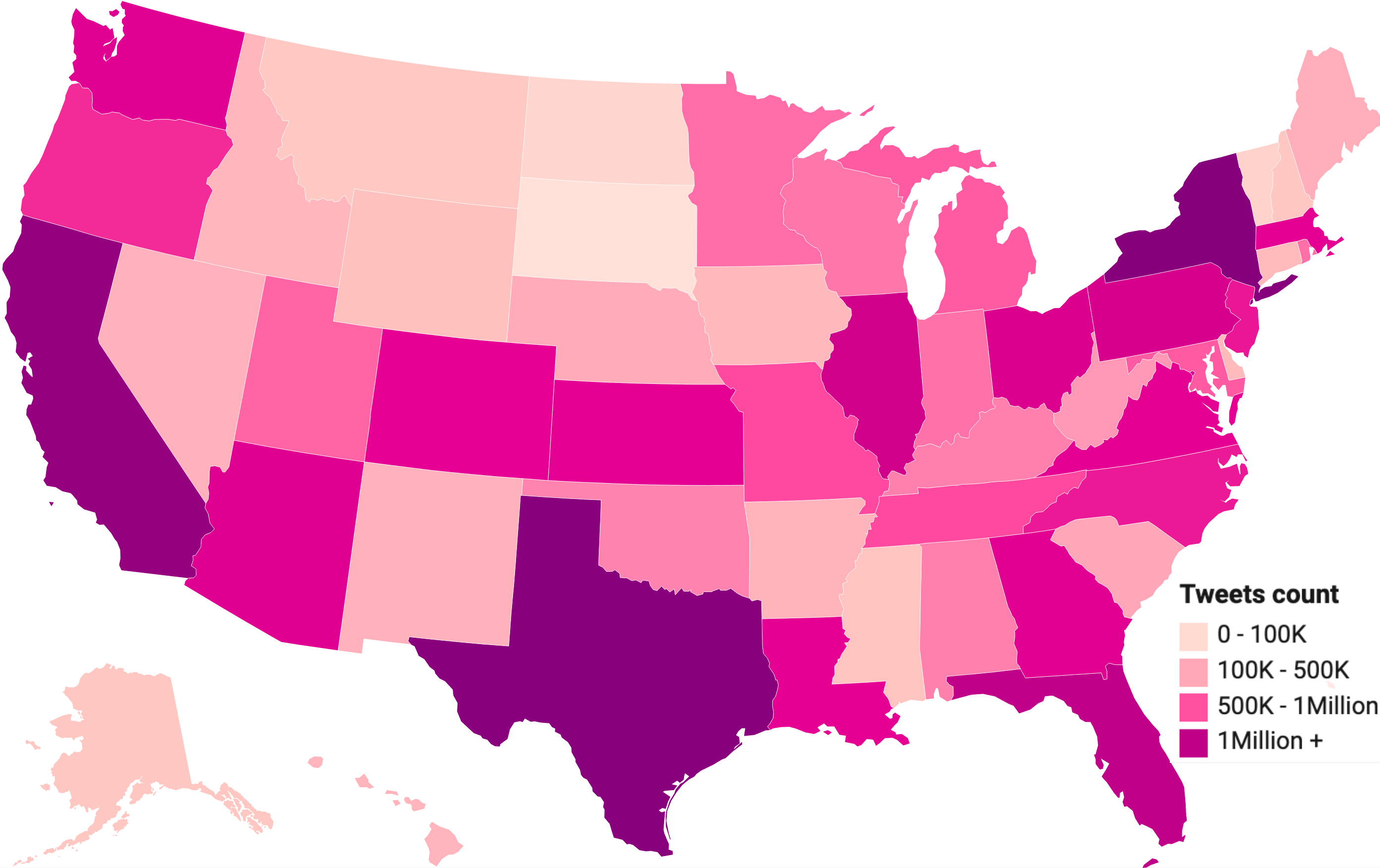}}
\subfigure[USA: Most-mentioned locations]{\label{fig:usa_most_mentioned_map}\includegraphics[width=0.32\textwidth]{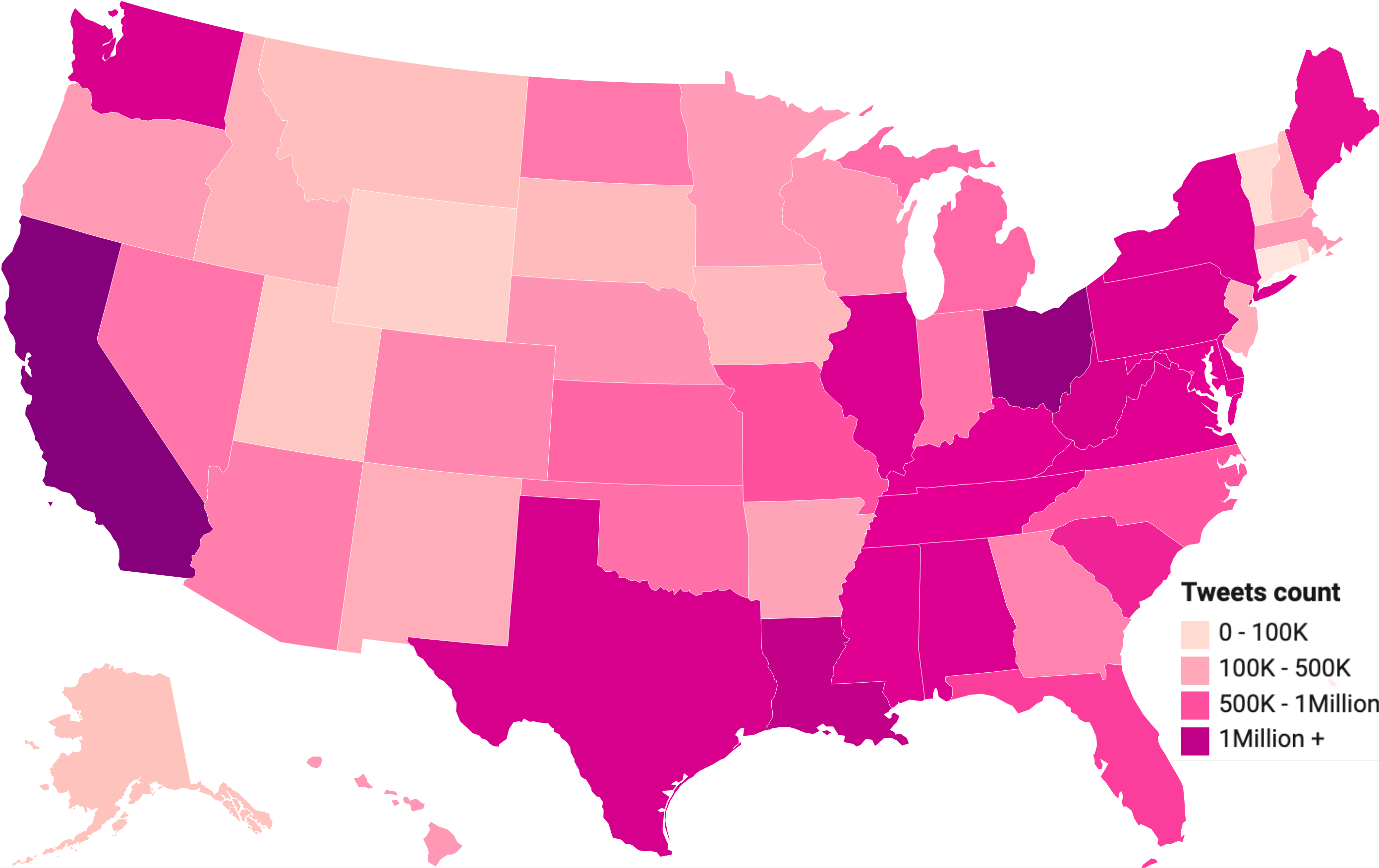}}
\end{minipage}

\hfill        
\begin{minipage}[b]{1\textwidth}
\centering
\subfigure[Italy: Organic geotagged tweets]{\label{fig:italy_geo_map}\includegraphics[width=0.32\textwidth]{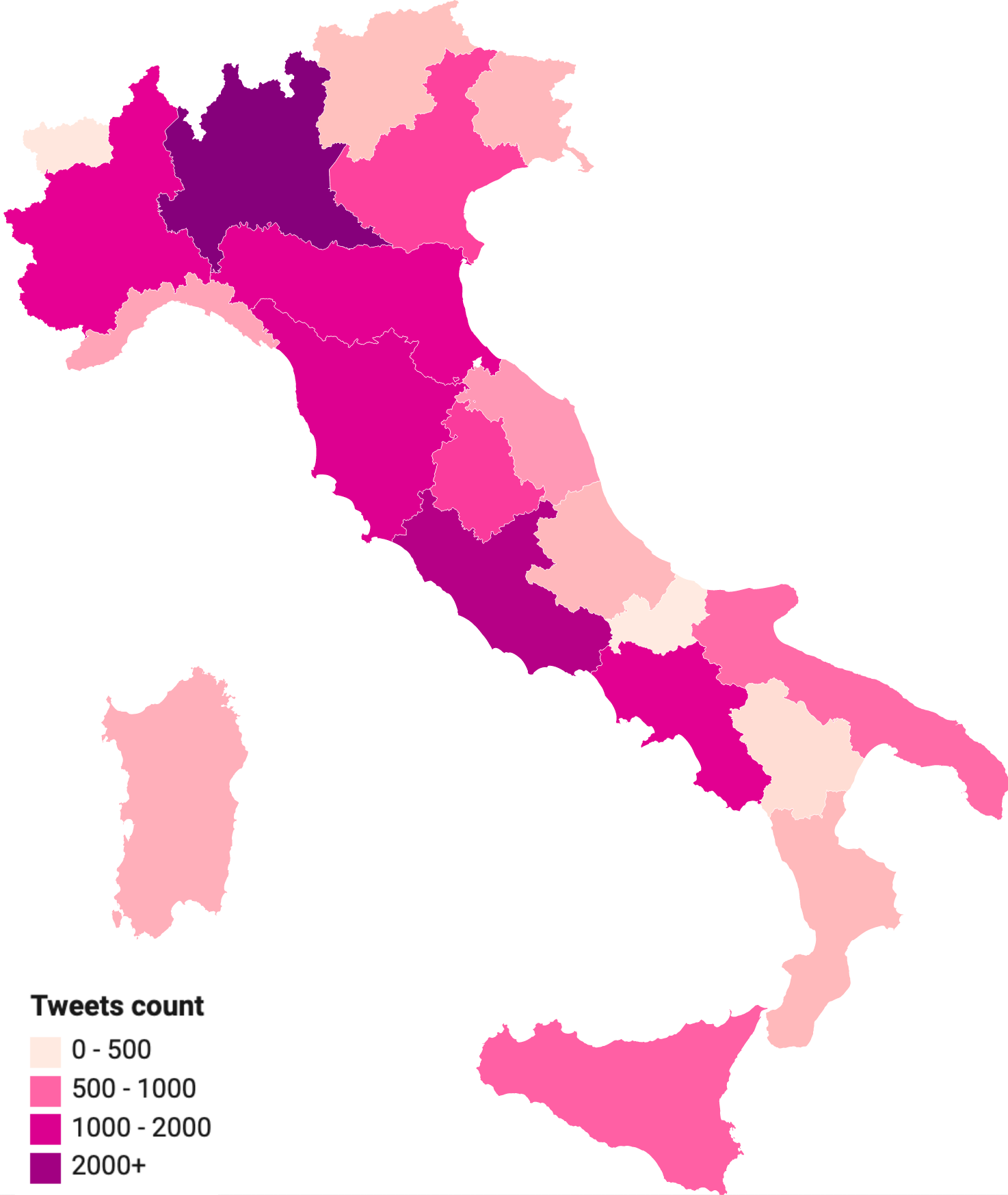}}
\subfigure[Italy: User distribution]{\label{fig:italy_users_map}\includegraphics[width=0.32\textwidth]{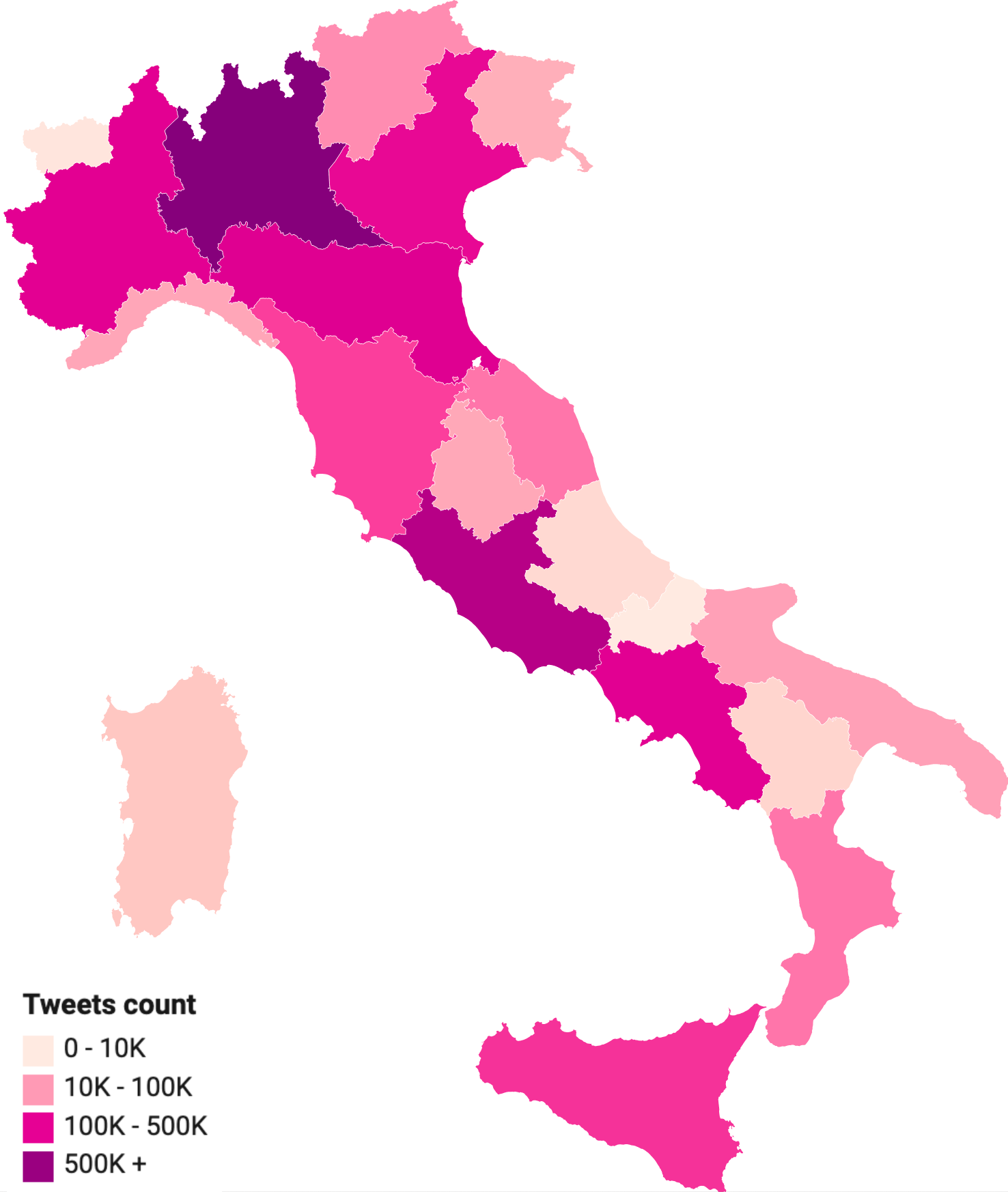}}
\subfigure[Italy: Most-mentioned locations]{\label{fig:italy_most_mentioned_map}\includegraphics[width=0.32\textwidth]{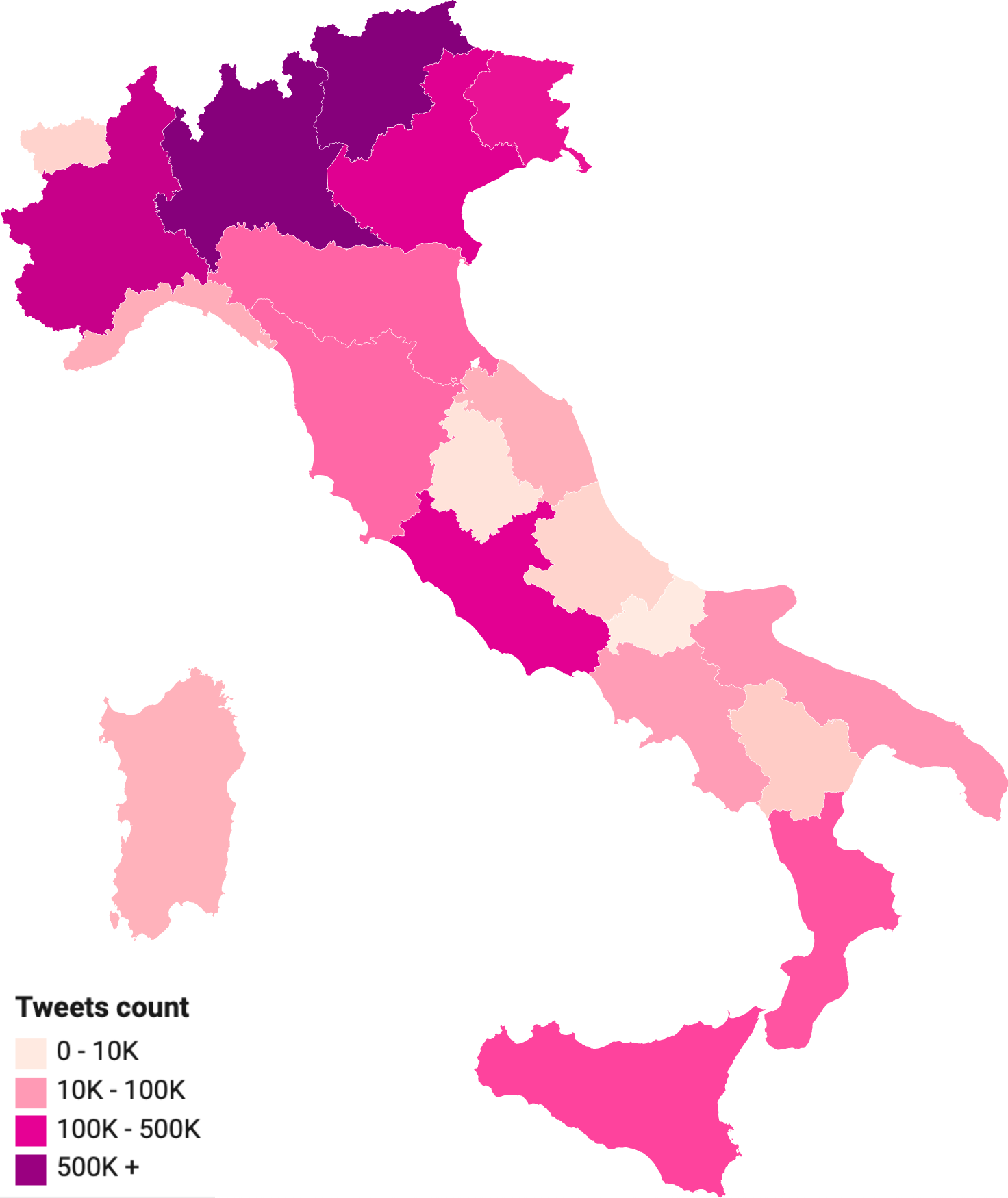}}
\end{minipage}

\hfill        
\begin{minipage}[b]{1\textwidth}
\centering
\subfigure[France: Organic geotagged tweets]{\label{fig:france_geo_map}\includegraphics[width=0.32\textwidth]{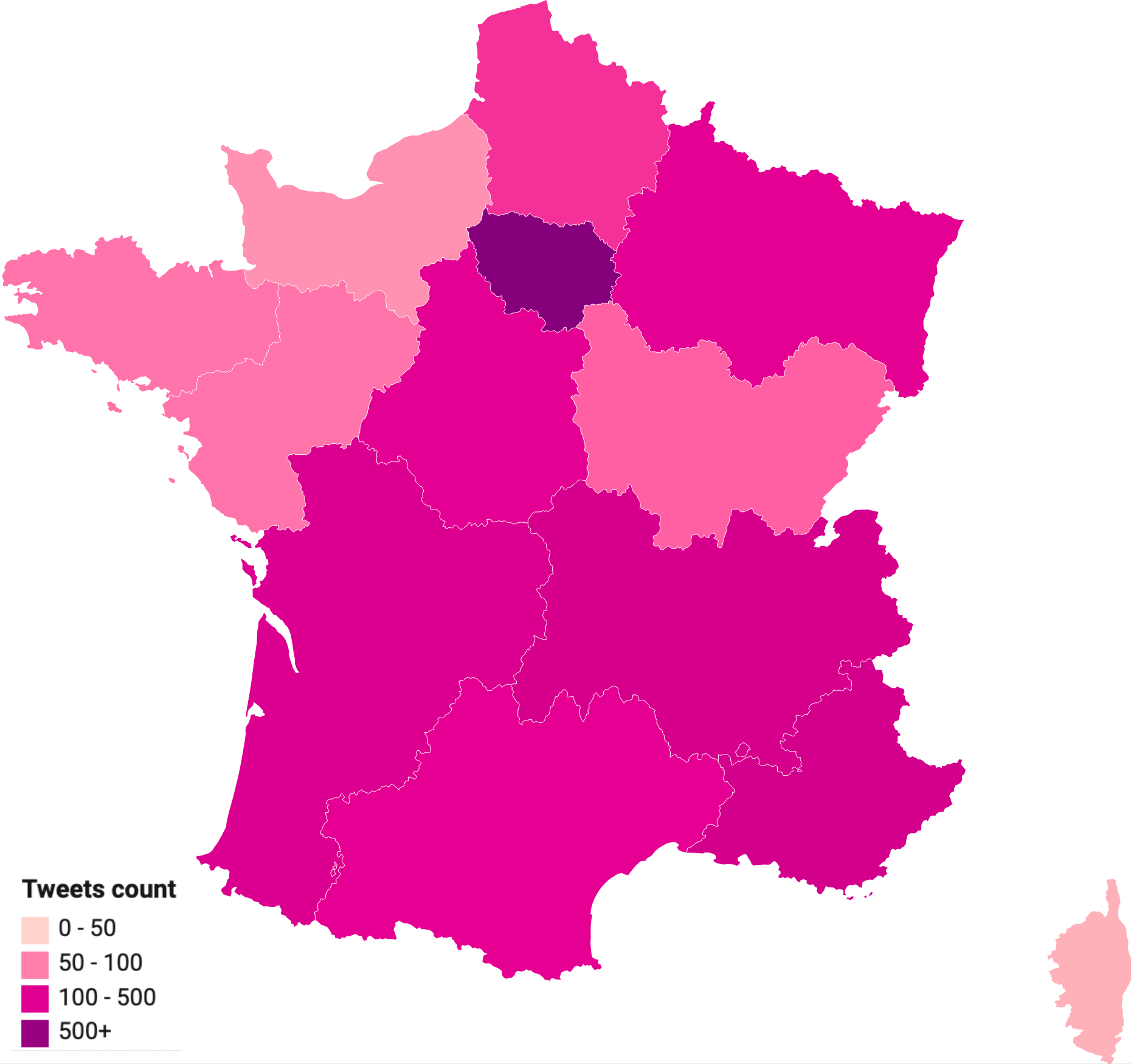}}
\subfigure[France: User distribution]{\label{fig:france_users_map}\includegraphics[width=0.32\textwidth]{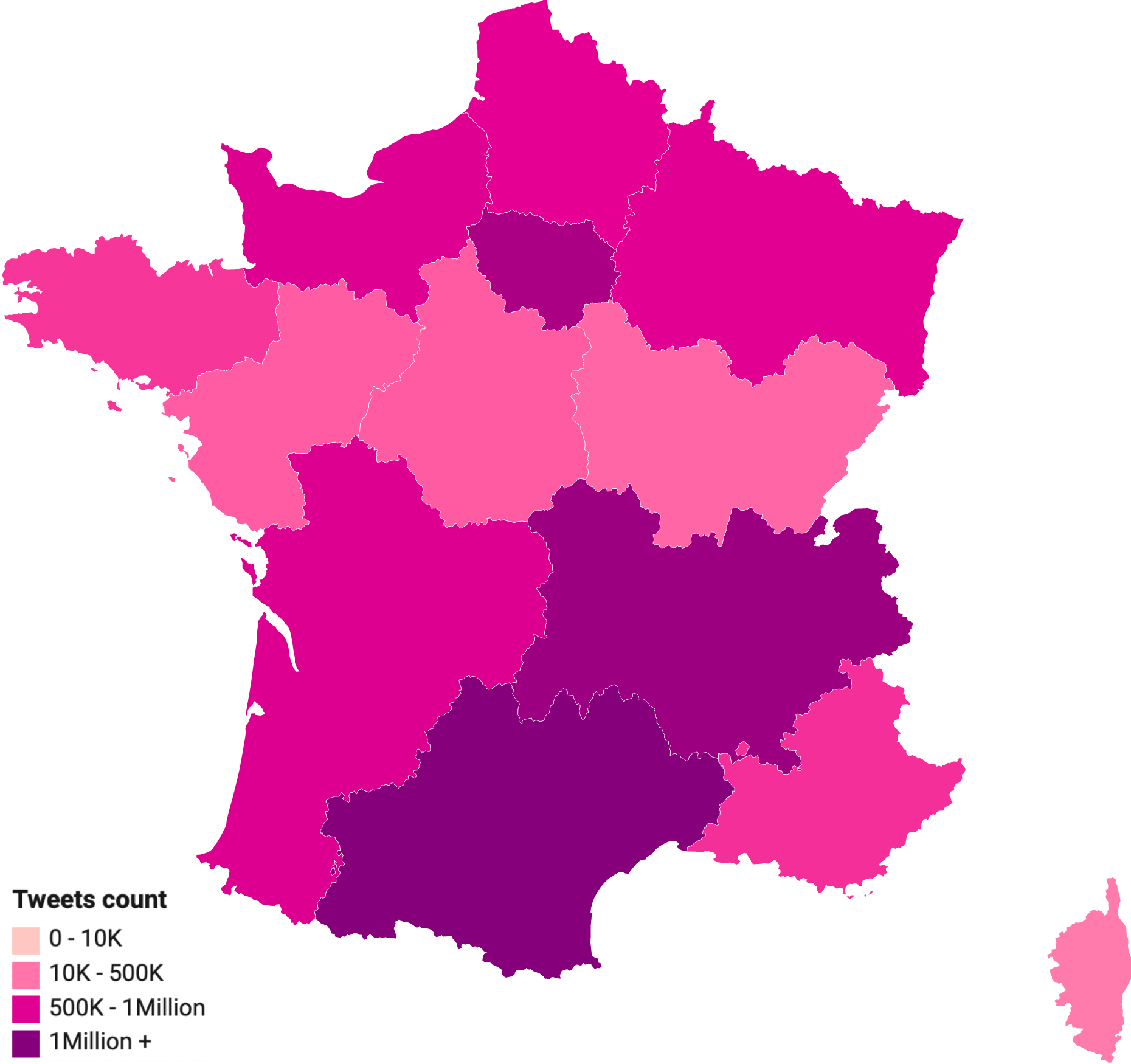}}
\subfigure[France: Most-mentioned locations]{\label{fig:france_most_mentioned_map}\includegraphics[width=0.32\textwidth]{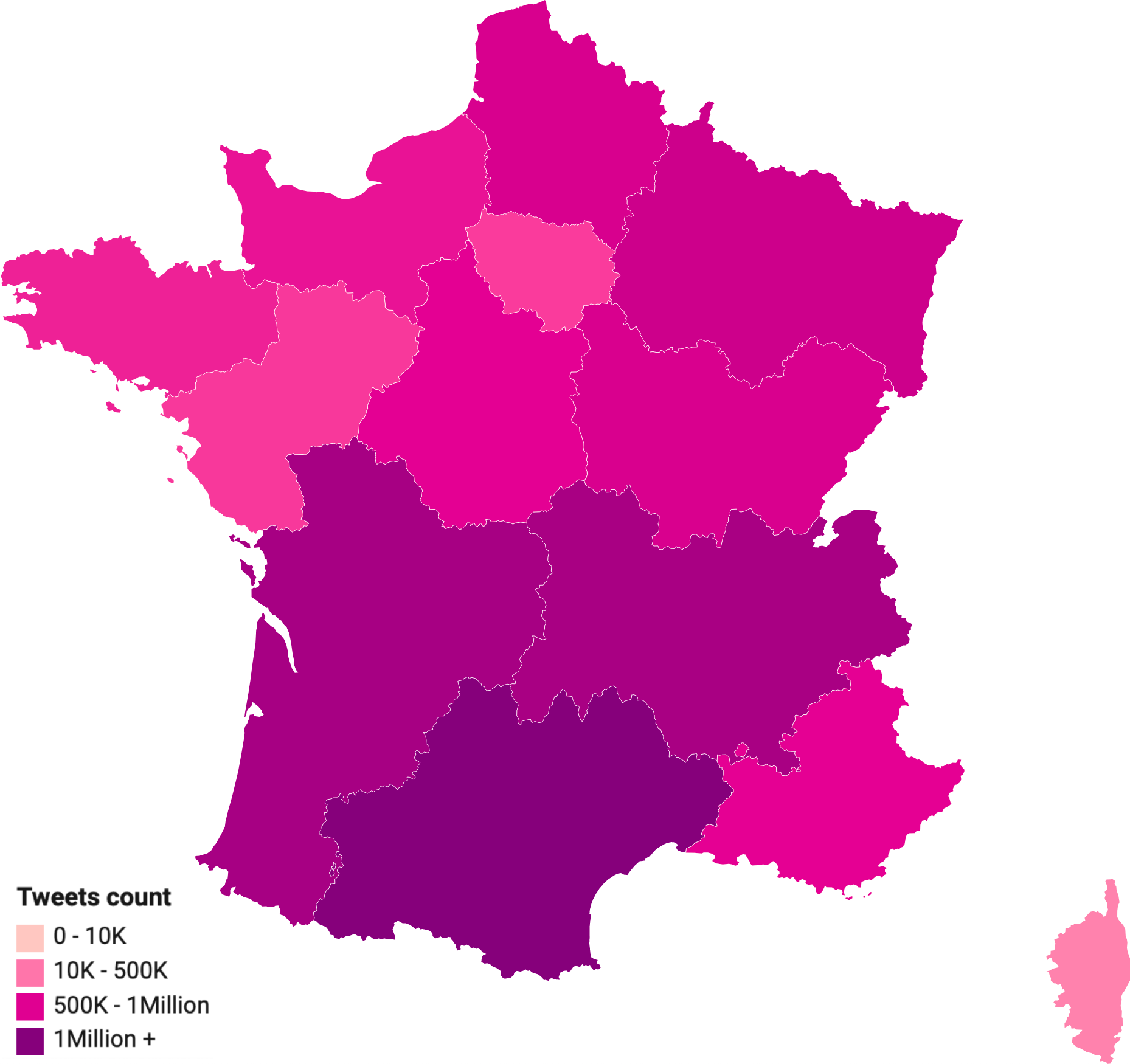}}
\end{minipage}

\hfill        
\begin{minipage}[b]{1\textwidth}
\centering
\subfigure[Spain: Organic geotagged tweets]{\label{fig:spain_geo_map}\includegraphics[width=0.32\textwidth]{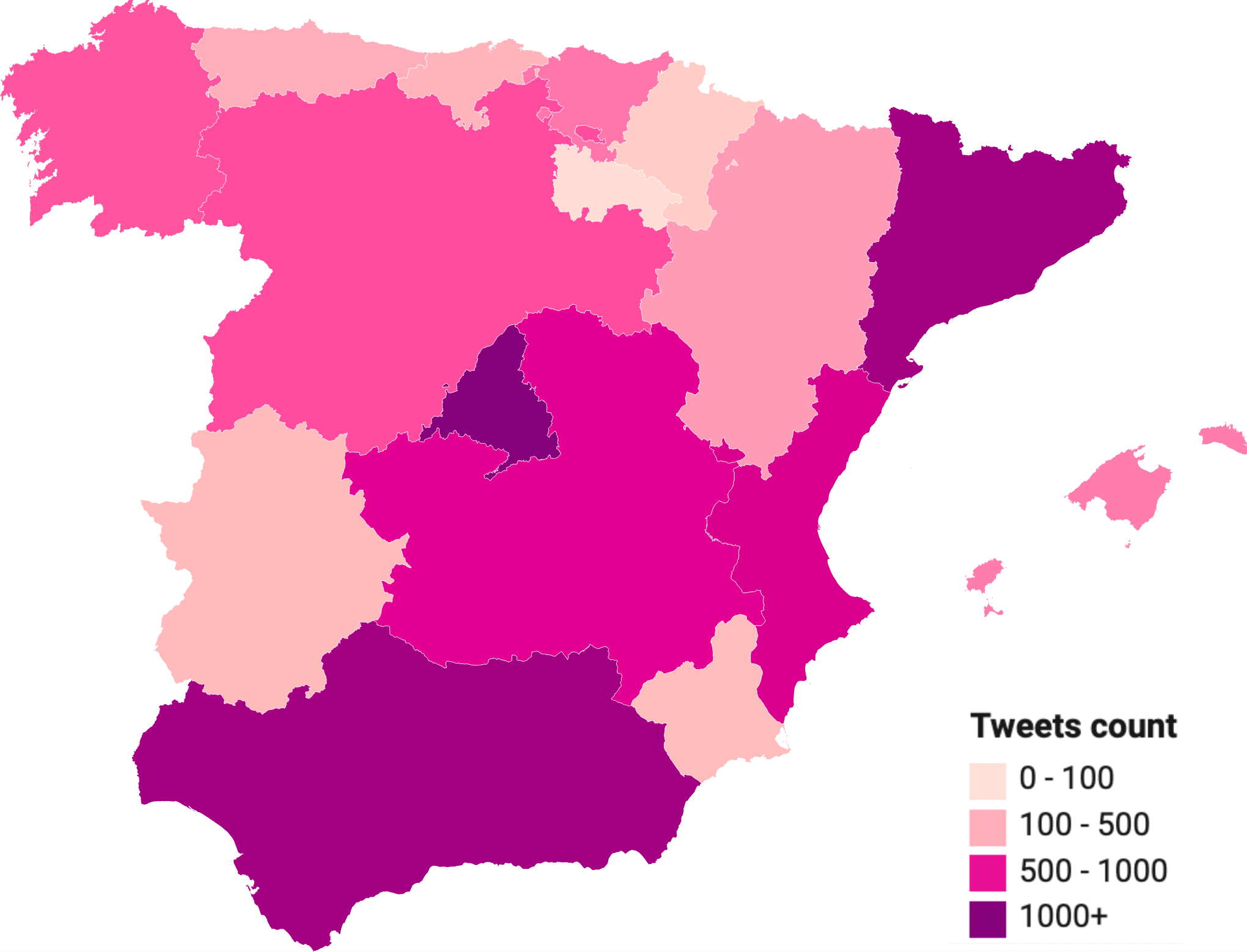}}
\subfigure[Spain: User distribution]{\label{fig:spain_users_map}\includegraphics[width=0.32\textwidth]{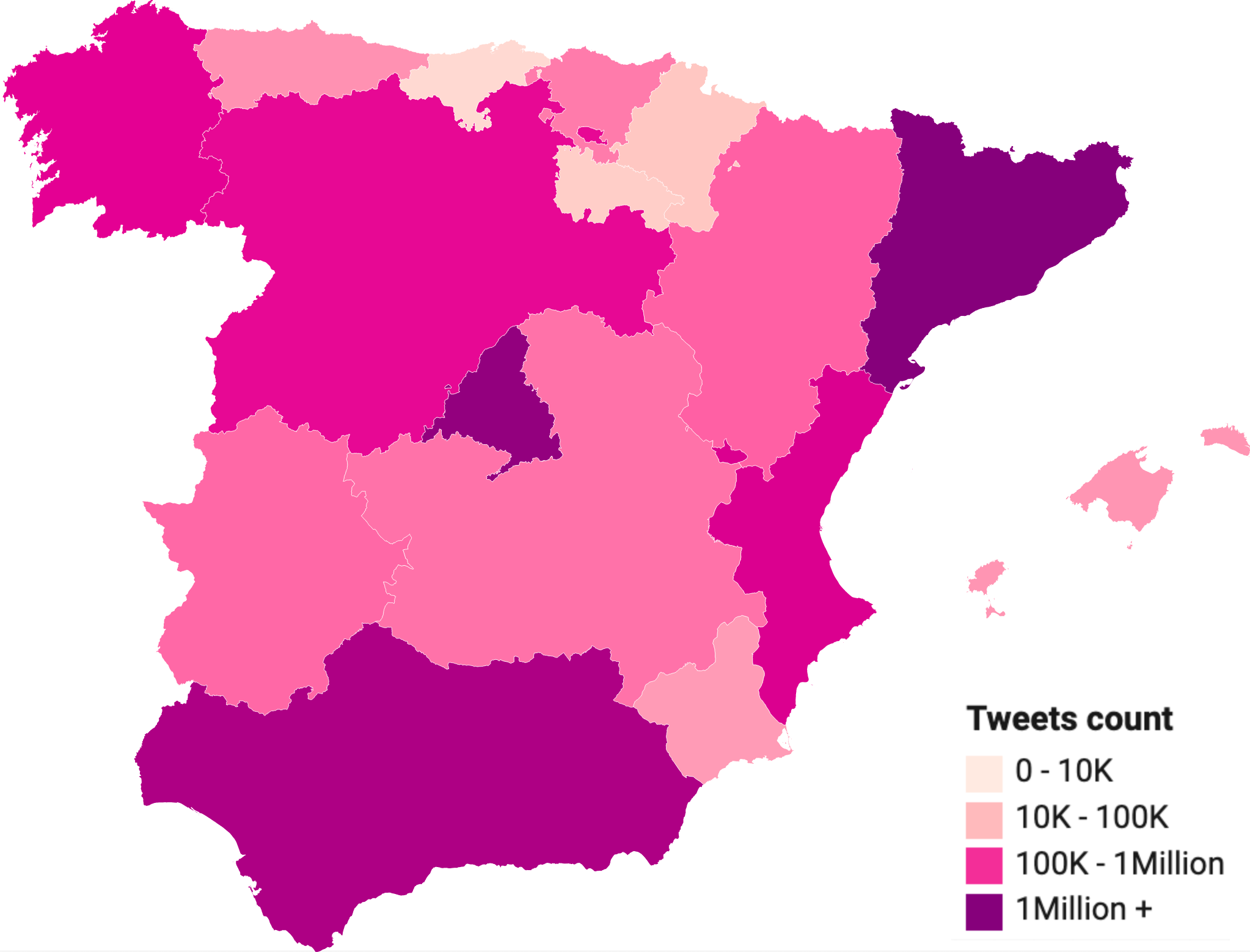}}
\subfigure[Spain: Most-mentioned locations]{\label{fig:spain_most_mentioned_map}\includegraphics[width=0.32\textwidth]{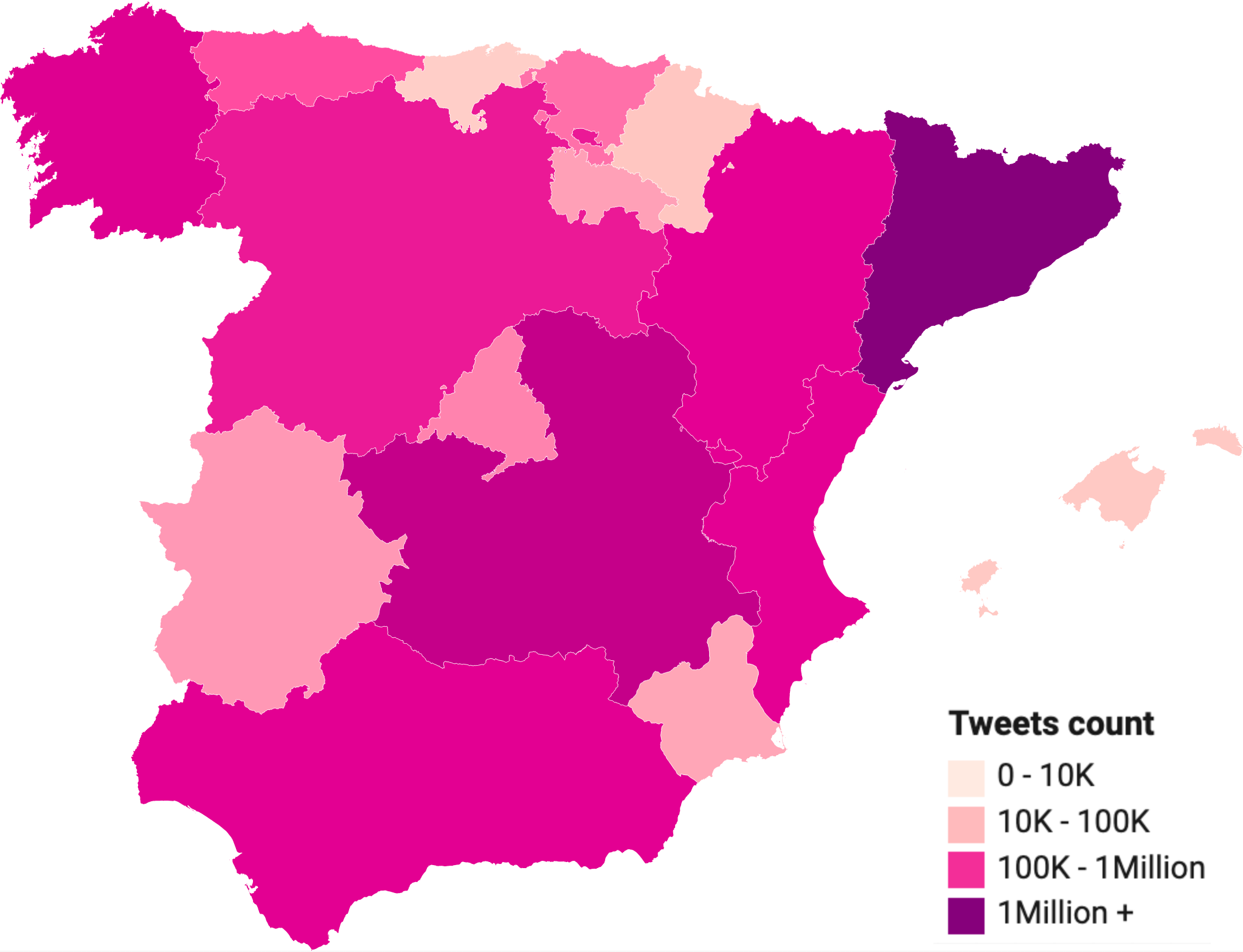}}
\end{minipage}
\caption{Country-specific maps for the United States, Italy, France, and Spain with geographic distribution of geotagged tweets, users, and most-mentioned locations}
\label{fig:four_countries}
\end{figure}

\begin{table}[h!]
\begin{minipage}[b]{0.32\textwidth}
\centering
\caption{Tweets with geolocation}
\begin{tabular}{lr} 
 \toprule
 Location field  & \# of tweets \\ 
 \midrule
 Geo-coordinates & 378,772\\
 Place & 5,495,431 \\
 User location & 297,148,292 \\
 Tweet text &  452,933,900 \\
 \bottomrule
\end{tabular}
\label{tbl:tweets_geo}
\end{minipage}
\hfill%
\begin{minipage}[b]{0.27\textwidth}
\centering
\caption{Countries breakdown}
\begin{tabular}{cl} 
 \toprule
 Countries & \# of tweets \\ 
 \midrule
 9 & $>10$M \\
 40 & $>1$M \\
 28 & $>500$K \\
 66 & $>100$K \\
 \bottomrule
\end{tabular}
\label{tbl:country_vol_bucket}
\end{minipage}
\hfill%
\begin{minipage}[b]{0.27\textwidth}
\centering
\caption{Cities breakdown}
\begin{tabular}{cl} 
 \toprule
 Cities & \# of tweets \\ 
 \midrule
 12 & $>1$M \\
 20 & $>500$K \\
 166 & $>100$K \\
 209 & $>50$K \\
 \bottomrule
\end{tabular}
\label{tbl:cities_vol_bucket}
\end{minipage}
\end{table}

The four countries which are severely hit by the COVID-19 pandemic include the United States, Italy, France, and Spain. To determine the country-level geographic coverage of tweets, in Figure~\ref{fig:four_countries}, we show the geographical distribution of tweets with geo-coordinates, tweets for which the location information is extracted from the user location, and most-mentioned locations which are extracted from the tweet content.



\smallskip
\smallskip

\noindent{\bf User-related statistics}\\
Table~\ref{tbl:users_unique_stats} shows various details of the users in the dataset. The dataset contains 43,463,225 unique users. Of all, 136,448 users tweeted at least one geotagged tweet and 1,355,032 users tagged at least one tweet with the \textit{place} information. Around 22,792,120 users have valid \textit{user location} value. Table~\ref{tbl:verified_users_stats} shows the details of verified users in the dataset. In total, the dataset contains 209,372 verified users and among them 1,853 users posted at least one geotagged tweet and 21,216 users with \textit{place} information, and 165,495 with valid user location value.

\begin{table}[h!]
\begin{minipage}[b]{0.465\textwidth}
\centering
\caption{User-related statistics}
\resizebox{\textwidth}{!}{
\begin{tabular}{lr}
 \toprule
Users (unique)  & \# of tweets \\ 
 \midrule
Unique users in the dataset & 43,463,225 \\
Users with at least one geo tweet & 136,448  \\
Users with at least one place tweet & 1,355,032  \\
Users with valid user location value & 22,792,120 \\
 \bottomrule
\end{tabular}
}
\label{tbl:users_unique_stats}
\end{minipage}
\hfill %
\begin{minipage}[b]{0.53\textwidth}
\centering
\caption{Statistics related to verified users}
\resizebox{\textwidth}{!}{
\begin{tabular}{lr} 
 \toprule
Verified users (unique)  & \# of tweets \\ 
 \midrule
Verified users in the dataset & 209,372 \\
Verified users with at least one geo tweet & 1,853 \\
Verified users with at least one place tweet & 21,216   \\
Verified users with valid user location value & 165,495   \\
 \bottomrule
\end{tabular}
}
\label{tbl:verified_users_stats}
\end{minipage}
\end{table}

\smallskip
\smallskip

\begin{figure}
\begin{minipage}[b]{0.46\textwidth}
\centering
\subfigure[Language frequency (log scale)]{
\label{fig:langs_all}
\includegraphics[width=\textwidth]{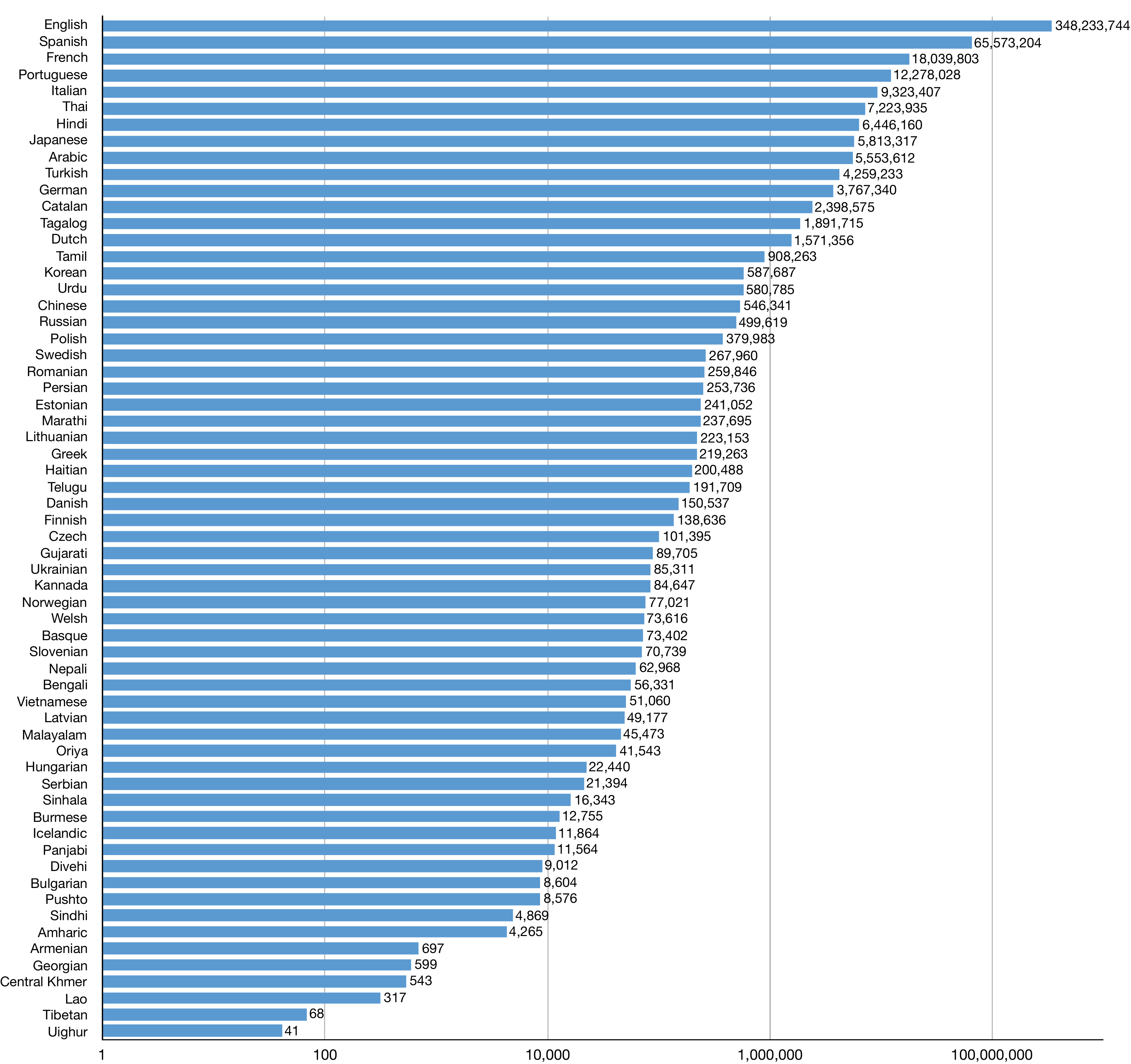}
}
\end{minipage}
\hfill        
\begin{minipage}[b]{0.52\textwidth}
\centering
\subfigure[Top 1-20 languages daily distribution]{
\label{fig:top_1_20_lang_dist}
\includegraphics[width=\textwidth]{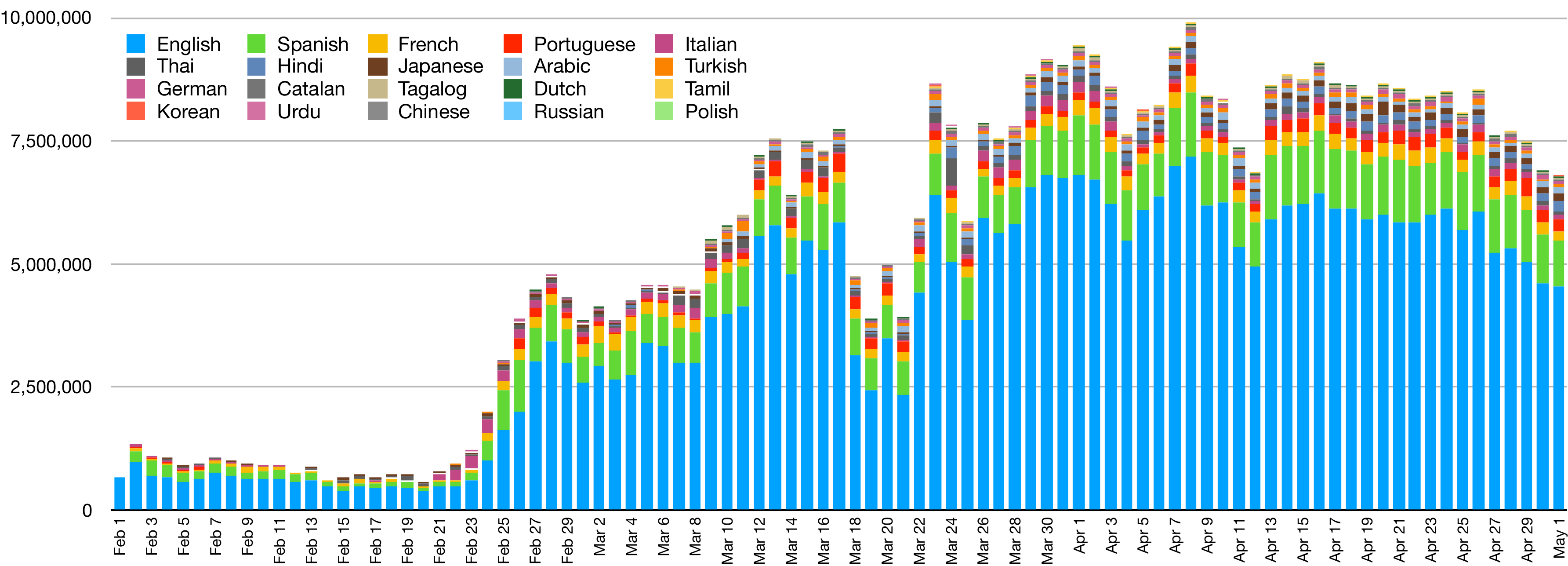}
}
\subfigure[Top 21-40 languages daily distribution]{
\label{fig:top_21_40_lang_dist}
\includegraphics[width=\textwidth]{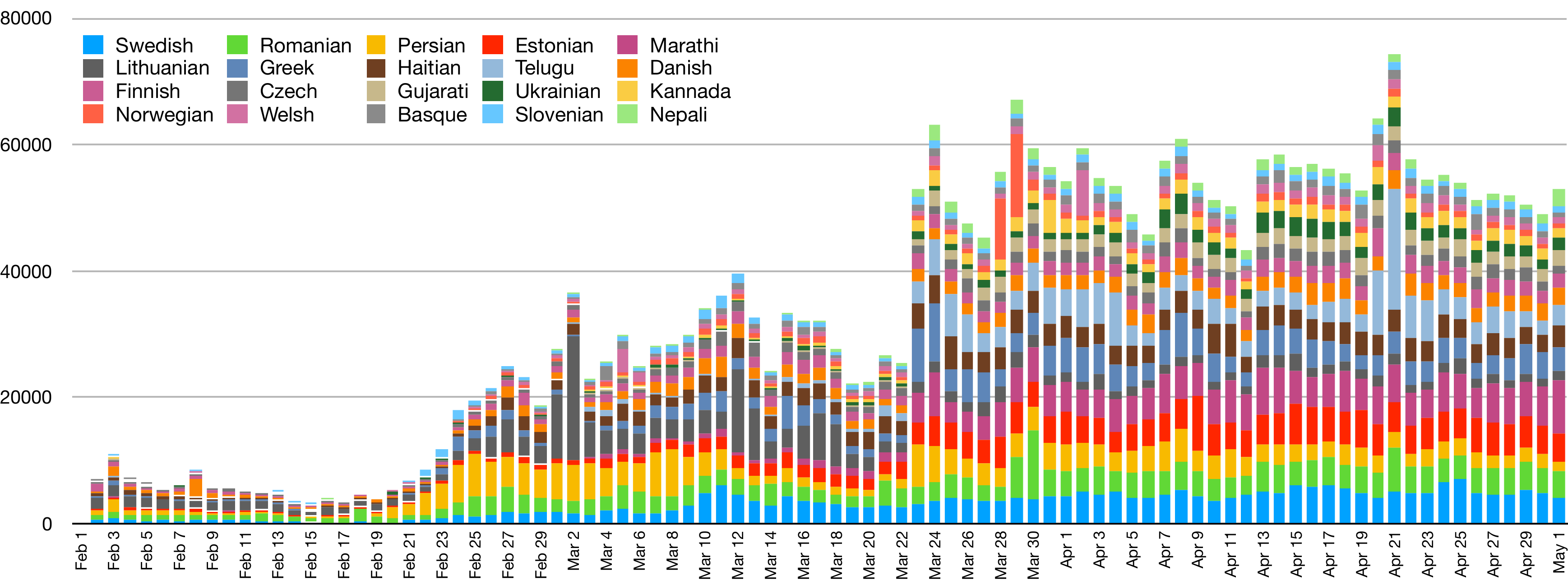}
}
\end{minipage}\vspace{-0.2cm}
\caption{All languages covered in the dataset and their distributions}\vspace{-0.5cm}
\end{figure}

\noindent{\bf Language-related statistics}\\
In total, the dataset covers 62 international languages. Figure~\ref{fig:langs_all} shows all the languages and the corresponding number of tweets on a log scale. The English language dominates with 348 million tweets and the second and third largest languages are Spanish and French, respectively. There are around 12 million tweets for which the language is undetermined. Figure~\ref{fig:top_1_20_lang_dist} and \ref{fig:top_21_40_lang_dist} show the daily distribution of top-40 languages.

\vspace{-0.4cm}\section{Research Implications}\vspace{-0.2cm}
\label{sec:apps}
Datasets are fundamental building blocks for both core and applied research. The recent widespread adoption of social media platforms, especially during emergency events, provides timely access to citizen-generated data, which can be useful for several tasks including event forecasting, surveillance, and response. The GeoCoV19 dataset presented in this paper is the largest COVID-19 related Twitter dataset available to date with broad coverage of geographical regions as well as languages, which we believe can foster multidimensional research in many application areas including the ones listed below.  

\begin{itemize}
    \item {\bf Understanding knowledge gaps:} Emergency events such as COVID-19 pandemic bring uncertainty, which raises questions that the general public asks through social media. Identification of such questions can help authorities to understand knowledge gaps among masses and address them as quickly as possible. One can look at country-specific tweets in GeoCoV19 dataset to identify knowledge gaps.
    \item {\bf Disease forecast and surveillance:} Early detection of disease outbreaks can prevent further spread and loss of lives. Automatic identification of messages where people report signs and symptoms of a disease can be used to identify hot spots where authorities can prioritize. Besides the textual content, images, and videos associated with the tweets in our dataset can also be used to assess the public's adherence to policies such as social distancing and mask usage.
    \item {\bf Identifying urgent needs of individuals:} At the onset of emergency situations like COVID-19, when governments have no choice other than closing businesses and imposing lockdowns, low-income communities, especially in developing countries, suffer. The lack of financial resources could even affect their daily meal intake, leaving them without sufficient nutrition. Such communities may have several other types of urgent needs, which can be identified through social media data. This dataset can be explored and further analyzed to develop automatic approaches to identify such unanticipated needs of the general public. Such automatic methods can then be employed during a future event.
    \item {\bf Tracking misinformation and fake news:} During critical events, rapid identification of misinformation, fake news, and false rumors is of utmost importance for authorities to turn them down as quickly as possible. Identification of such content on social media is even more important due to the fact that such information spreads quickly on social media platforms compared to other new media channels. There were several times when fake information related to COVID-19 treatment associated with WHO appeared on social media. The dataset covers many such fake news stories, which can be helpful in developing misinformation and fake news detection systems. In addition to the textual content, analyzing the visual content associated with tweets (i.e., images, videos, animated GIFs) can provide further insights about the flow of misinformation (e.g., memes) and help assess public's reaction to such cases.
    \item {\bf Understanding public reactions and sentiment:} Social media platforms are a great source to learn about issues and difficulties that the general public is facing and their reactions to government response during emergency events. COVID-19 poses several unanticipated challenges for the general public, businesses, as well as authorities, which can be learned from this dataset. Especially joint analysis of textual and visual content in our dataset can afford the ability to develop models for monitoring mental health consequences of the pandemic on the public.
\end{itemize}

\vspace{-0.7cm}\section{Conclusion}\vspace{-0.2cm}
\label{sec:conclusion}
In this paper, we present GeoCoV19, a large-scale Twitter dataset related to the ongoing COVID-19 pandemic. The dataset has been collected over a period of 90 days from February 1 to May 1, 2020 and consists of more than 524 million multilingual tweets. As the geolocation information is essential for many tasks such as disease tracking and surveillance, we employed a gazetteer-based approach to extract toponyms from user location and tweet content to derive their geolocation information using the Nominatim (Open Street Maps) data at different geolocation granularity levels. In terms of geographical coverage, the dataset spans over 218 countries and 47K cities in the world. The tweets in the dataset are from more than 43 million Twitter users, including around 209K verified accounts. These users posted tweets in 62 different languages. We posit that GeoCoV19 affords the ability to develop computational models to have a better understanding of how societies are collectively coping with the unprecedented COVID-19 pandemic. Moreover, the dataset can inform the development of AI-based systems to forecast disease outbreaks as well as provide surveillance for authorities to act timely, learn about knowledge gaps and urgent needs of the general public, identify unanticipated issues, and tackle misinformation and fake news, among others. 


\bibliographystyle{abbrv}
\bibliography{sigproc}

\end{document}